\documentclass[twocolumn,trackchanges,twocolappendix]{aastex63}

\usepackage[nameinlink]{cleveref}
\usepackage{ragged2e}
\usepackage[]{natbib}
\setcitestyle{comma}
\usepackage[utf8]{inputenc}
\usepackage[T1]{fontenc}

\makeatletter
\newcommand*{\linktocite}[2]{%
  \hyper@natlinkstart{#1}#2\hyper@natlinkend}
\makeatother

\newcommand{\Sec}[1]{Section~\ref{#1}}

\newcommand{\Fig}[1]{Figure~\ref{#1}}

\def\K{\; {\rm K}}

\journalinfo{Accepted at ApJL}

\shorttitle{Effect of the interior on the atmospheric circulation of (ultra)-hot Jupiters}
\shortauthors{Komacek, Gao, Thorngren, May, \& Tan}
\graphicspath{{./}{figs/}}

\begin{document}

\title{The effect of interior heat flux on the atmospheric circulation of hot and ultra-hot Jupiters}

\correspondingauthor{Thaddeus D. Komacek}
\email{tkomacek@umd.edu}
\author[0000-0002-9258-5311]{Thaddeus D. Komacek}
\affiliation{Department of Astronomy, University of Maryland, College Park, MD 20742, USA}
\author[0000-0002-8518-9601]{Peter Gao}
\affiliation{Earth and Planets Laboratory, Carnegie Institution for Science, 5241 Broad Branch Road, NW, Washington, DC 20015, USA}
\author[0000-0002-5113-8558]{Daniel P. Thorngren}
\affiliation{Institute for Research on Exoplanets (iREx), Universit\'{e} de Montr\'{e}al, Canada}
\author[0000-0002-2739-1465]{Erin M. May}
\affiliation{Johns Hopkins APL, 11100 Johns Hopkins Road, Laurel, MD 20723, USA}
\author[0000-0003-2278-6932]{Xianyu Tan}
\affiliation{Atmospheric, Oceanic and Planetary Physics, Department of Physics, University of Oxford, OX1 3PU, UK}




\begin{abstract}
Many hot and ultra-hot Jupiters have inflated radii, implying that their interiors retain significant entropy from formation. These hot interiors lead to an enhanced internal heat flux that impinges upon the atmosphere from below. In this work, we study the effect of this hot interior on the atmospheric circulation and thermal structure of hot and ultra-hot Jupiters. To do so, we incorporate the population-level predictions from evolutionary models of hot and ultra-hot Jupiters as input for a suite of General Circulation Models (GCMs) of their atmospheric circulation with varying semi-major axis and surface gravity. We conduct simulations with and without a hot interior, and find that there are significant local differences in temperature of up to hundreds of Kelvin and in wind speeds of hundreds of m s$^{-1}$ or more across the observable atmosphere. These differences persist throughout the parameter regime studied, and are dependent on surface gravity through the impact on photosphere pressure. These results imply that the internal evolution and atmospheric thermal structure and dynamics of hot and ultra-hot Jupiters are coupled. As a result, a joint approach including both evolutionary models and GCMs may be required to make robust predictions for the atmospheric circulation of hot and ultra-hot Jupiters.
\end{abstract}



\section{Introduction} 
\label{sec:intro}
Many transiting hot and ultra-hot Jupiters have radii larger than expected from standard evolutionary models of irradiated gas giants \citep{fortney_2009,Baraffe:2014}. This implies that they have a high-entropy internal convective zone and a hot central temperature \citep{Ginzburg:2015} to sustain these bloated radii. There is an observed dependence of planetary radius on irradiation \citep{Demory11,Laughlin_2011}, which implies that the mechanism which sustains the high interior entropy of hot and ultra-hot Jupiters is deposition of a fraction of incident stellar flux as heat in the interior  \citep{Thorngren:2018,Sarkis:2021aa,Thorngren:2021aa}. This further indicates that the atmosphere has an important role in regulating interior heat deposition, either through current generation in the atmosphere and resulting Ohmic dissipation \citep{baty10,Perna10,Rauscher_2013,Wu:2013,Rogers:2014,ginz16} or effective downward heat transport by the atmospheric circulation itself \citep{Guillot_2002,showman_2002,Youdin10,Tremblin:2017}.  

The high interior entropy of hot and ultra-hot Jupiters necessitates that they have a high internal heat flux consistent with their radius evolution \citep{Thorngren:2019aa}. However, many previous hot Jupiter GCMs (e.g., \citealp{Showmanetal_2009,Heng:2011a,Rauscher_2012,Komacek:2017}) did not consider the impact of this enhanced internal heat flux on atmospheric circulation. More recent work has studied the coupled nature of the interior evolution and atmospheric dynamics of hot and ultra-hot Jupiters, revising the standard model of their atmospheric circulation (for recent reviews covering the atmospheric circulation of hot and ultra-hot Jupiters, see \citealp{Showman:2020rev,Zhang:2020rev,Fortney:2021aa}). This includes GCMs conducted with deep and/or long-timescale integrations \citep{Mayne:2017,Carone:2019aa,Mendonca:2020aa}, 
which can assess whether the deep layers could converge to an adiabat at depth \citep{Sainsbury-Martinez:2019aa,Sainsbury-Martinez:2021us,Schneider:2022uu}. Alternatively, recent work by \cite{Lian:2022wc} has incorporated interior thermal perturbations that mimic convection in simulations of irradiated giant planets, finding that waves triggered by the internal forcing pattern can in turn affect the shallow atmospheric circulation. 

In this work, we consistently incorporate predictions from the  population-level interior evolution models of \cite{Thorngren:2018} and \cite{Thorngren:2019aa} as bottom boundary conditions in 3D GCMs of the atmospheric circulation of hot and ultra-hot Jupiters. We conduct a large suite of simulations across a range of semi-major axis (i.e., irradiation) and surface gravity, both with a consistently hot interior and with the standard assumption of a relatively cold interior. We find that there are significant differences in the predicted thermal structure and wind pattern between the cases with and without a hot interior, and that these differences persist across the parameter regime of semi-major axis and surface gravity considered. 

This work is outlined as follows. We describe our GCM setup in \Sec{sec:methods}, including the parameter sweep of our GCM suite covering interior assumptions, semi-major axis, and surface gravity. We present results for the influence of the interior on atmospheric circulation in \Sec{sec:results}. Lastly, we discuss our results and limitations and summarize conclusions in \Sec{sec:conc}. 
\section{GCM Setup}
\label{sec:methods}
\subsection{Dynamical core and radiative transfer}
In this work, we conduct simulations of hot and ultra-hot Jupiters with the {\tt MITgcm} \citep{Adcroft:2004} coupled to the two-stream {\tt TWOSTR} package \citep{Kylling:1995} of the {\tt DISORT} radiative transfer code \citep{Stamnes:2027}. We use the ultra-hot Jupiter version of the {\tt MITgcm} that has been modified to take into account the thermodynamic impact of hydrogen dissociation and recombination on atmospheric dynamics \citep{Tan:2019aa,Mansfield:2020aa,May:2021ab,Komacek:2022tf}. This GCM solves the primitive equations of meteorology on a cubed-sphere grid, including the thermodynamic impact of hydrogen dissociation and recombination in the energy conservation equation along with the resulting spatial variation in specific gas constant and heat capacity as described in \cite{Tan:2019aa}. Following \cite{Tan:2019aa,Mansfield:2020aa}, and \cite{May:2021ab}, we include the thermodynamic effects of hydrogen dissociation and recombination assuming an atmosphere comprised purely of hydrogen, and do not include recent updates by \cite{Komacek:2022tf} to incorporate inert helium. We use a double-gray radiative transfer scheme with one visible and one infrared opacity band, with the same pressure-dependent opacity formulation as \cite{Tan:2019aa}. Double-gray radiative transfer schemes have been used in many recent studies of exoplanet atmospheric dynamics (e.g., \citealp{Dietrick:2020aa,May:2020vr,Mendonca:2020aa,Roman:2021wl,Beltz:2022aa,Komacek:2022tf}), and allow for the exploration of a broad parameter space (see \Sec{sec:parametersweep}) while retaining a physically realistic treatment of radiation. 
\subsection{Parameter sweep}
\label{sec:parametersweep}

In order to ascertain the impact of interior evolution on the atmospheric circulation, we consider two separate assumptions for the bottom boundary in the radiative transfer scheme, resulting in two separate suites of GCM simulations. The first suite of GCM simulations (which we term ``fixed flux'') imposes a fixed net flux at the bottom boundary of the GCM \citep{Showmanetal_2009,Rauscher_2012,Kataria:2014}, corresponding to an intrinsic temperature of $T_\mathrm{int} = 100~\mathrm{K}$, which is comparable to previous GCM studies (e.g., \citealp{Showmanetal_2009,Heng:2011a,Kataria:2014}). Models in the second suite (which we term ``hot interior'') 
use a bottom boundary condition with a specified total (\textit{not} net) upward heat flux (e.g., \citealp{Dobbs-Dixon:2013,Mendonca:2020aa}), which effectively relaxes the temperature at the bottom boundary (at 100 bars, which we term $T_{100}$) toward the desired value informed by the evolutionary model \citep{May:2021ab,Komacek:2022tf}.
The hot interior suite with a prescribed bottom temperature mimics vigorous deep convection that connects the deep domain of the GCM to the interior profile. 

For each case of bottom boundary condition, we conduct a grid of models varying the semi-major axis around a Solar-type host star (0.0125 au, 0.015 au, 0.0175 au, 0.02 au, 0.025 au, 0.03 au, 0.04 au)\footnote{These correspond to full-redistribution, zero albedo equilibrium temperatures of 2494.8 K, 2277.4 K, 2108.5 K, 1972.3 K, 1764.1 K, 1610.4 K, 1394.6 K.} and the surface (1-bar) gravity (4 m s$^{-2}$ and 10 m s$^{-2}$). 
We use corresponding 100-bar temperatures in the hot interior cases from the combined interior-atmosphere simulations of \cite{Thorngren:2019aa,Gao:2020aa}\footnote{In the case with a surface gravity of 4 m s$^{-2}$ the 100-bar temperatures for each semi-major axis are 4708.4 K, 4654.1 K, 4593.3 K, 4533.3 K, 4405.4 K, 4265.6 K, 3864.8 K; with a surface gravity of 10 m s$^{-2}$ they are 4432.8 K, 4405.7 K, 4359.5 K, 4313.2 K, 4191.8 K, 4047.3 K, 3562.1 K.}. As in \cite{Thorngren:2019aa}, we consistently vary the radius with surface gravity and semi-major axis in order to take into account the dependence of the deposited heating that causes radius inflation on irradiation \citep{Thorngren:2018,Sarkis:2021aa}. In order to study the effects of interior heat flux, semi-major axis (i.e., instellation), and surface gravity on the atmospheric dynamics, as in \cite{Tan:2019aa} we keep the rotation period fixed at $P_\mathrm{rot} = 2.43~\mathrm{days}$. As a result, though these simulations assume tidal locking, the rotation period does not consistently vary with irradiation\footnote{Note that for our assumed stellar and planetary parameters, the rotation period would vary from 0.51 - 2.92 days with increasing semi-major axis from $0.0125-0.04$ au.} in order to keep the Coriolis parameter fixed and isolate the combined impacts of irradiation and interior heat flux. In summary, we conduct fourteen cases for each interior assumption over seven values of irradiation and two values of surface gravity, resulting in a total of twenty-eight GCMs in our combined suite. 

\subsection{Numerical details}
\label{sec:numerics}

As in \cite{Komacek:2022tf}, all simulations contain a weak Rayleigh drag throughout the domain with a characteristic drag timescale $\tau_\mathrm{drag,uniform} = 10^7~\mathrm{s}$. Similarly to \cite{Liu:2013} and \cite{Komacek:2015}, each simulation has a deep basal drag with a characteristic timescale that decreases linearly in pressure from $\tau_\mathrm{drag,basal} = \infty$ at 10 bars to a characteristic drag of $\tau_\mathrm{drag,basal} = 10^5~\mathrm{s}$ at the bottom of the domain (analogous to Equation 12 of \citealp{Komacek:2015}). This drag is applied in order to ensure that each simulation can reach an equilibrated state near the photosphere independent of initial conditions \citep{Liu:2013}, limiting any potential hysteresis \citep{Thrastarson:2010,Sergeev:2022ab}. The resulting specific drag force applied in the momentum equation at each individual pressure level is then $\mathcal{F}_\mathrm{drag} = -{\bf v}/ \mathrm{min}\left(\tau_\mathrm{drag,uniform},\tau_\mathrm{drag,basal}\right)$. We choose this drag formalism in order to assess the impacts of interior assumptions using a similar model setup to previous hot and ultra-hot Jupiter simulations with the {\tt MITgcm} (e.g., \citealp{Liu:2013,Komacek:2017}). In order to assess the influence of our frictional drag prescription on the results, we also conducted simulations without a deep bottom drag, which are discussed further in Appendix \ref{app:drag}.

Simulations are conducted with a horizontal resolution of C48, which approximately corresponds to $192$ longitudinal and $96$ latitudinal grid points, with 70 vertical levels evenly spaced in log-pressure from 100 bar to $10~\mu\mathrm{bar}$. Note that \cite{Carone:2019aa} demonstrated the influence of the assumed bottom boundary pressure on the shallower atmospheric dynamics. Due to the challenges of reaching an equilibrated state with a thick atmosphere, we leave such deep and long-timescale integrations with the {\tt MITgcm} to future work.
The dynamical timestep for each simulation is $10~\mathrm{s}$ and the radiative transfer timestep is $20~\mathrm{s}$. All simulations were continued until they reached an equilibrated state in domain-integrated kinetic energy, which corresponds to a run-time of 3,500 Earth days. The domain-integrated thermal energy is equilibrated for our hot interior cases, while the thermal energy in fixed flux cases is equilibrated only at $p \lesssim 20~\mathrm{bar}$, sufficient for comparing the effect of interior evolution on the circulation near the photosphere. All results presented in this manuscript are time-averages over the last 500 Earth days of each simulation.

\section{Results}
\label{sec:results}
\subsection{Effect of interior heat flux on atmospheric circulation}
\label{sec:onecase}
Before comparing the results from our full grid of simulations with varying irradiation and surface gravity in \Sec{sec:irradgrav}, here we first study the impact of interior assumptions on the hottest cases in our model grid: an ultra-hot Jupiter with a full-redistribution equilibrium temperature of $2495~\mathrm{K}$. 
\subsubsection{Temperature structure}
\begin{figure*}
    \centering
    \includegraphics[width=0.99\textwidth]{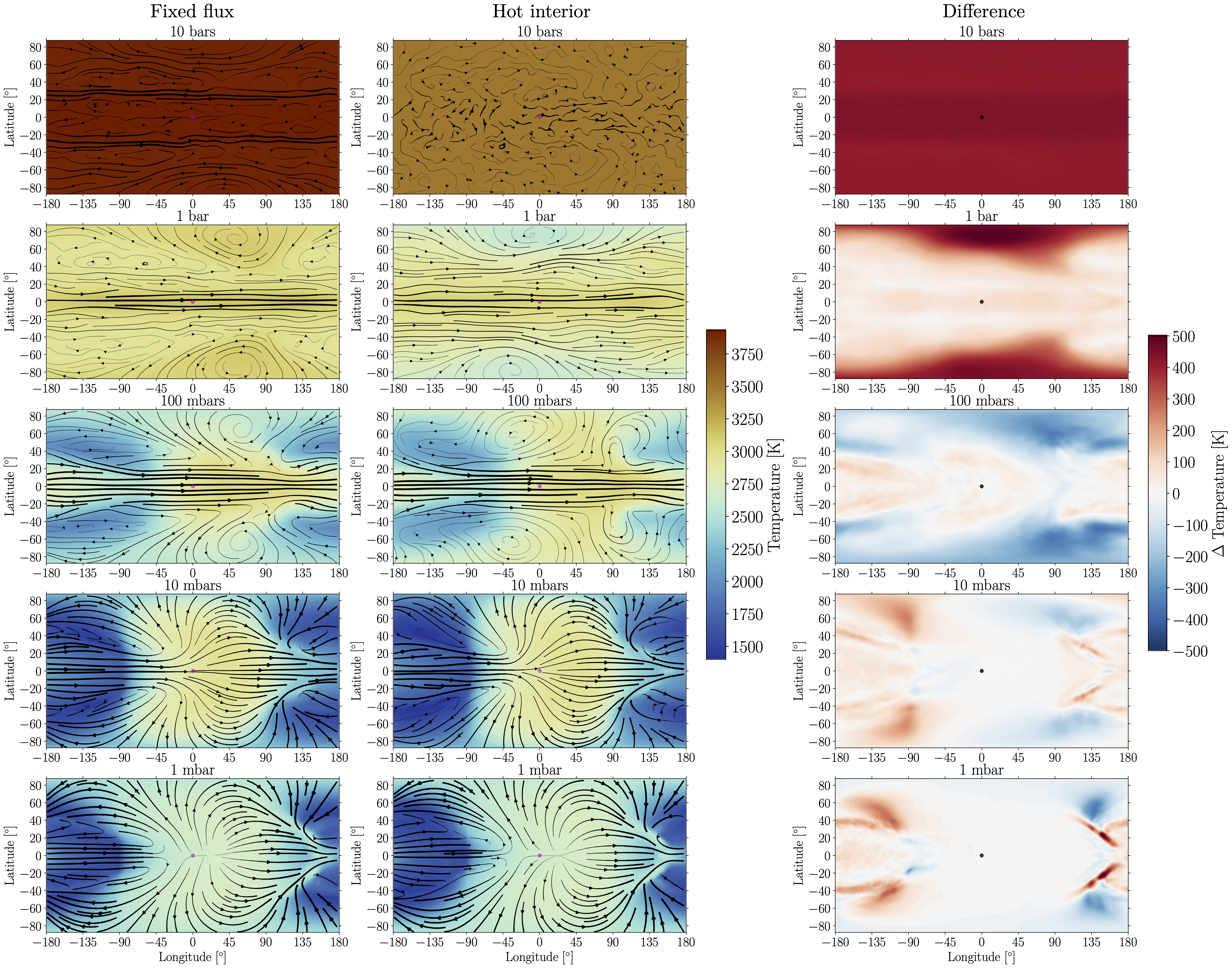}
    \caption{Temperature maps with overlaid wind streamlines plotted on isobars for the fixed flux and hot interior cases with $a = 0.0125~\mathrm{au}$ and $g = 4~\mathrm{m}~\mathrm{s}^{-2}$, along with the difference in temperature between the two cases (fixed flux temperature minus hot interior temperature). Streamline width scales with wind speed, but each panel has a separate streamline scale in order to discern qualitative differences in flow patterns. All panels in the two left-most columns and in the rightmost-column, respectively, share a color scale. The dot at the center of each map displays the substellar point. Due to the difference in interior assumptions, there are significant differences in the domain-wide temperature at 10 bars. Local temperature differences between the two simulations persist even to low pressures of $\sim 1~\mathrm{mbar}$.}
    \label{fig:tempmaps_4g}
\end{figure*}
\Fig{fig:tempmaps_4g} shows temperature maps on isobars for the hottest case with a surface gravity of $g = 4~\mathrm{m}~\mathrm{s}^{-2}$ for both the fixed flux (left) and hot interior (middle) assumptions, along with the local difference in temperature between the two simulations (right). There are significant local differences in temperature between the two simulations at all pressures shown. Notably, the fixed flux case has a globally hotter temperature at 10 bars than the hot interior case, while at $p < 1~\mathrm{bar}$ the hot interior case can be warmer than the fixed flux case locally.
The temperature differences between the two cases become more localized with decreasing pressure, and at pressures of 10-100 mbars are largest in the region of the equatorial jet and mid-latitude Rossby wave crests and troughs. The largest local temperature differences at the 1 mbar pressure level occur in regions of localized convergence that cause downwelling and adiabatic warming, described as local chevron-shaped features by \cite{Beltz:2022aa}\footnote{These are distinct from the global-scale chevron pattern described by \cite{Showman_Polvani_2011}, which is caused by the differential phase shift of Kelvin and Rossby waves that induces northwest-southeast wind vector tilts in the northern hemisphere and corresponding southwest-northeast tilts in the southern hemisphere.}. At pressures $< 1~\mathrm{mbar}$, the local temperature contrasts decrease with decreasing pressure, with maximum local temperature contrasts of $40~\mathrm{K}$ at $10~\mu\mathrm{bar}$ (not shown). 

\begin{figure}
    \centering
    \includegraphics[width=0.4\textwidth]{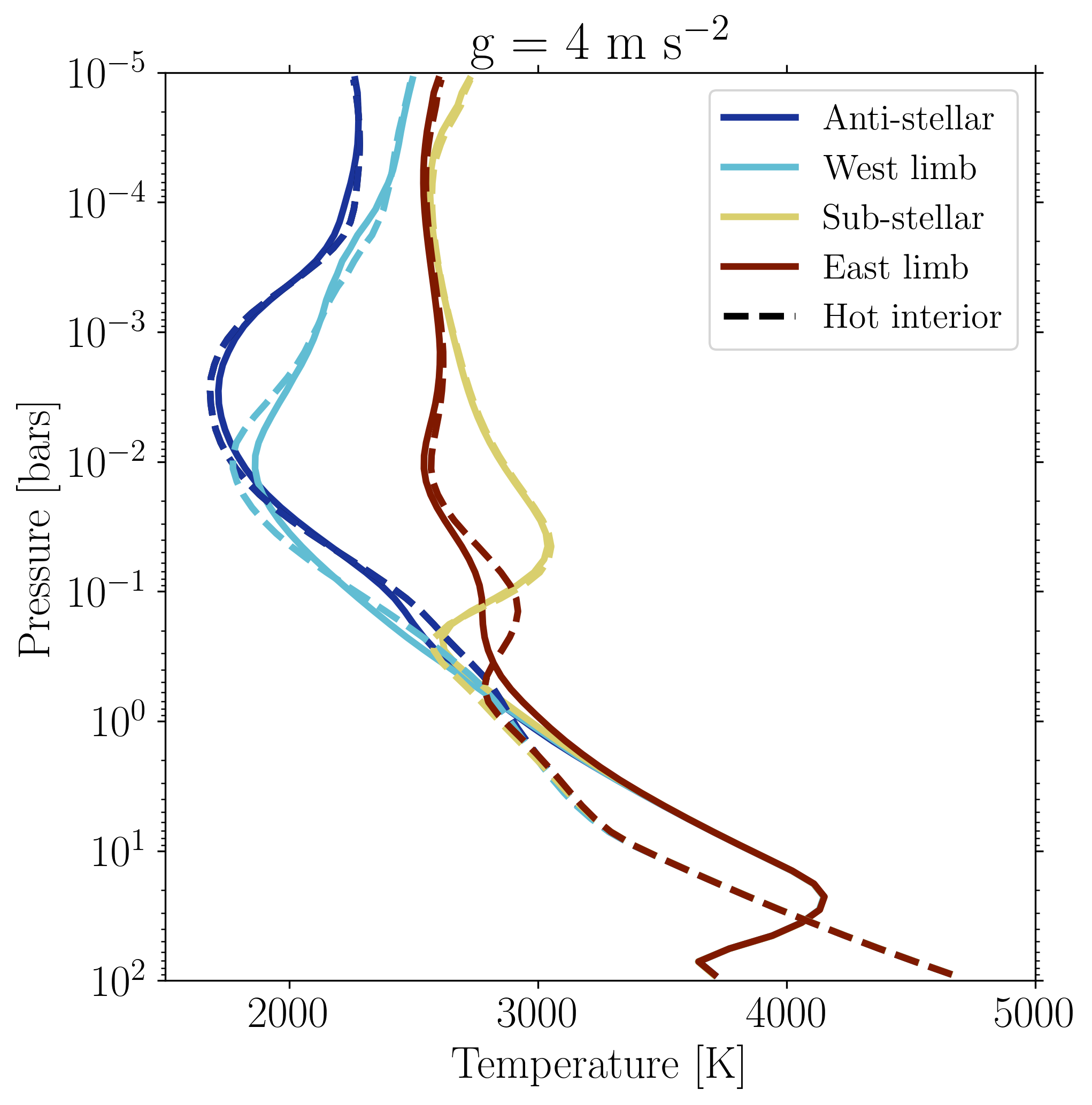}
    \includegraphics[width=0.4\textwidth]{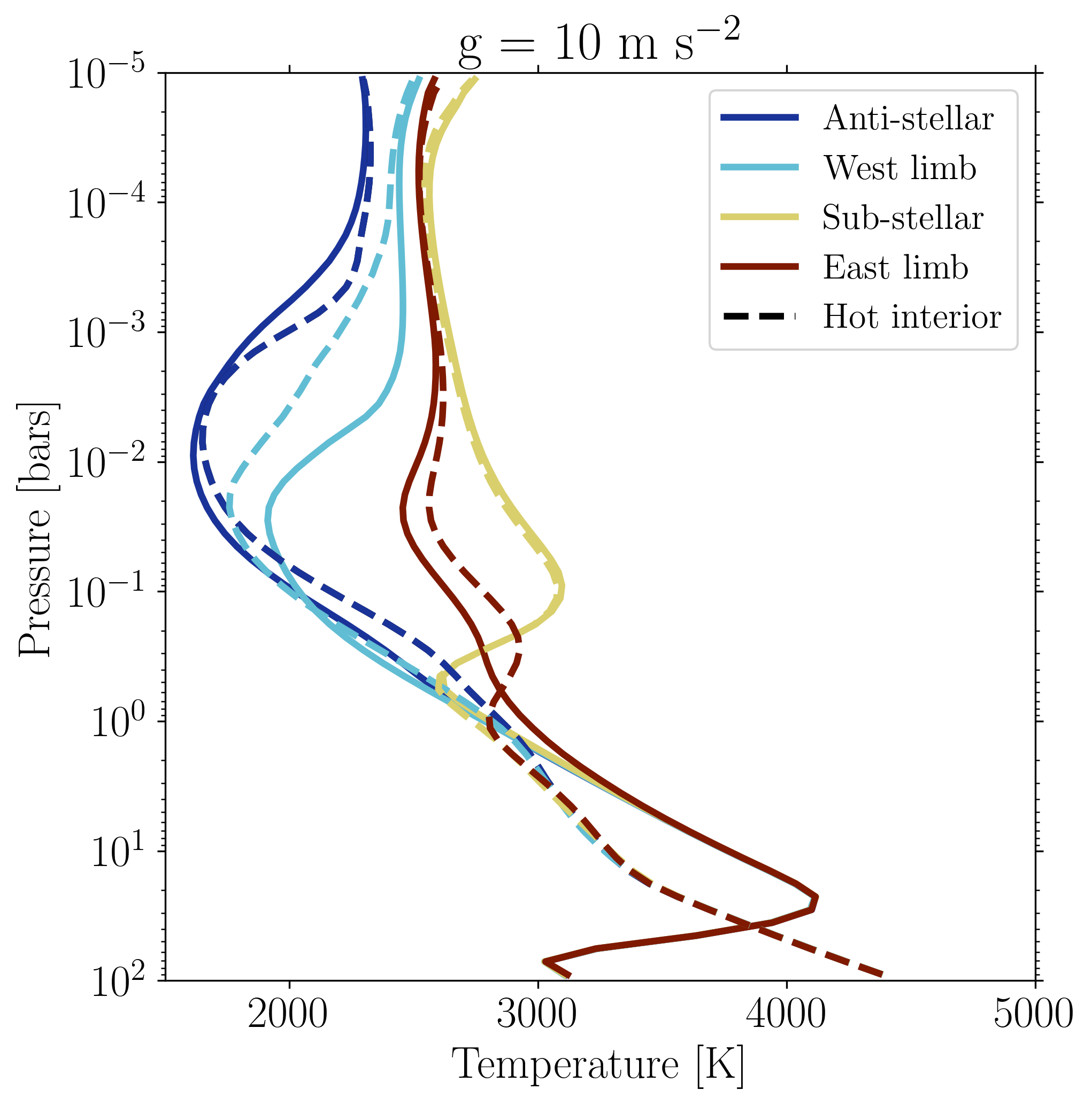}
    \caption{Meridional-mean temperature profiles at the anti-stellar point, west limb, sub-stellar point, and east limb in cases without (solid lines) and with (dashed lines) a hot interior. Both panels show results from simulations with $a = 0.0125~\mathrm{au}$, and the top panel shows the case of a surface gravity $g = 4~\mathrm{m}~\mathrm{s}^{-2}$ while the bottom panel shows $g = 10~ \mathrm{m}~\mathrm{s}^{-2}$. Temperature differences between the two cases are largest at pressures $\gtrsim 1~\mathrm{bar}$ due to the hotter 100-bar temperature 
    in the hot interior case. However, meridional-mean temperature differences between the two cases persist to the top of the simulation domain.}
    \label{fig:tempg_varyg}
\end{figure}
\Fig{fig:tempg_varyg} shows the temperature-pressure profiles averaged across latitude at the anti-stellar, west limb, sub-stellar, and eastern limb longitudes with both fixed flux (solid line) and hot interior (dashed line) assumptions for the cases of a surface gravity of $4~\mathrm{m}~\mathrm{s}^{-2}$ (as shown in \Fig{fig:tempmaps_4g}, top) and $10~\mathrm{m}~\mathrm{s}^{-2}$ (bottom). The most notable difference is the lack of a thermal inversion near the bottom of the domain in the hot interior case, which has been proposed to lead to cold trapping of condensible species at depth \citep{Parmentier16}. The thermal inversion near the bottom of the domain in our fixed flux cases occurs because the thermal structure in the deepest regions of the atmosphere is still evolving to an equilibrated state due to the long thermal adjustment timescales at depth \citep{Sainsbury-Martinez:2019aa,Schneider:2022aa,Schneider:2022uu}. Meanwhile, there is no appreciable time-evolution in the temperature profiles of both the fixed flux and hot interior cases at pressures $\le 10~\mathrm{bars}$. We find 
colder temperatures at the $1-10~\mathrm{bar}$ level in the hot interior case even though the temperature at $100~\mathrm{bars}$ is significantly warmer than in the fixed flux case. As for the local temperature maps in \Fig{fig:tempmaps_4g}, the meridional-mean temperature differences between the fixed flux and hot interior cases are significant throughout the atmosphere for both gravities. There are increased temperature contrasts between the two limbs in the hot interior case (largest in the higher gravity case), which could enhance the limb-to-limb inhomogeneity in cloudiness and transmission spectra \citep{Powell:2019aa,Espinoza:2021we}. Note that the deep temperature is cooler in the hot interior cases with a higher gravity. This is because evolutionary models \citep{Thorngren:2018} predict a cooler $T_{100}$ for higher gravity cases with a given internal heat flux due to the reduced thickness of higher-gravity atmospheres. 

\subsubsection{Wind speeds}
\label{sec:windspeeds}
\begin{figure*}
    \centering
    \includegraphics[width=0.99\textwidth]{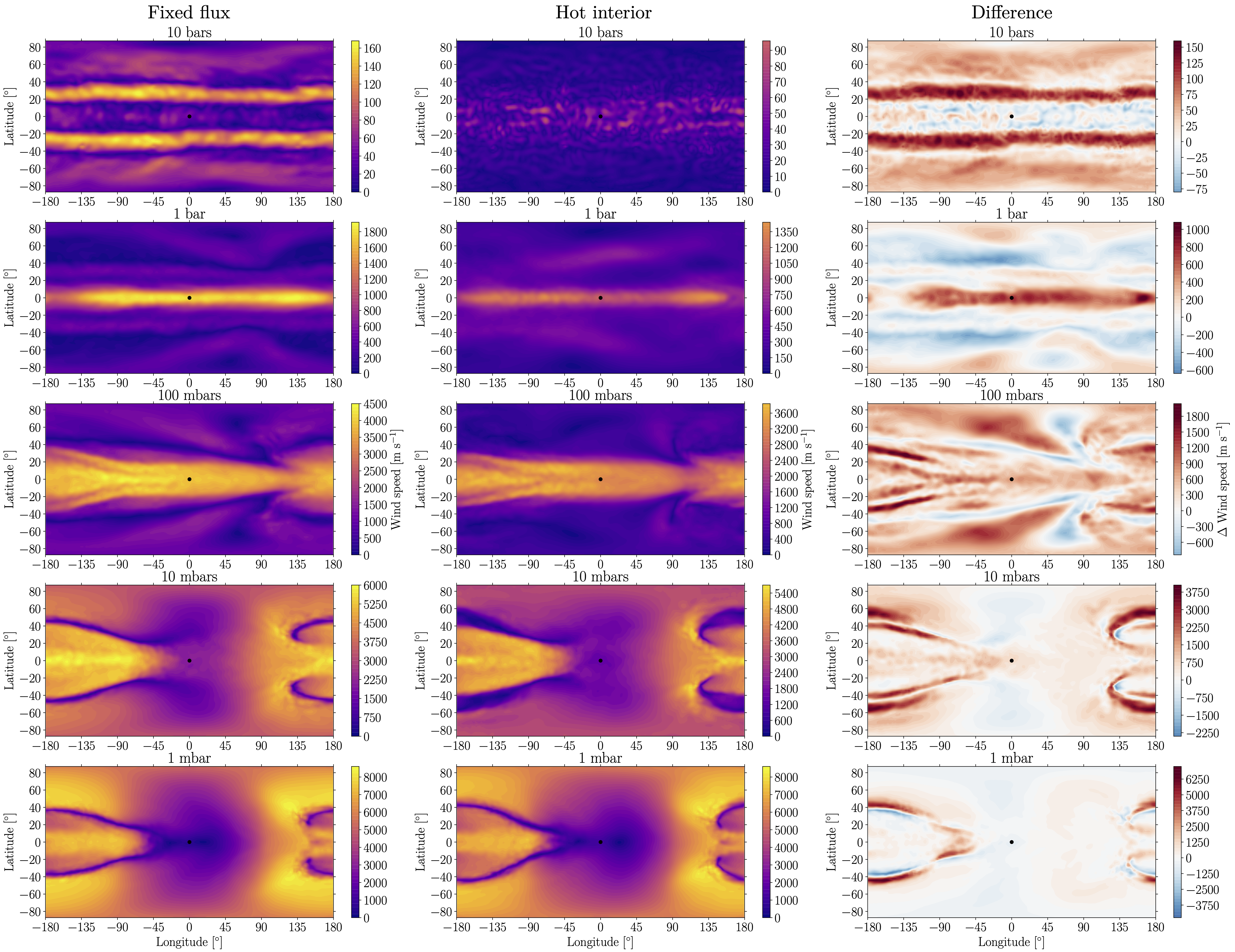}
    \caption{Horizontal wind speed on isobars for the fixed flux and hot interior cases with $a = 0.0125~\mathrm{au}$, along with the difference in wind speed between the two cases (fixed flux wind speed minus hot interior wind speed). For the fixed flux and hot interior columns, each row shares a color scale for wind speed. However, each plot in the difference column has a different color scale for the difference in wind speed. The fixed flux case has a stronger equatorial jet than the hot interior case, and local differences in wind speed on the order of km s$^{-1}$ occur at pressures $\le 1~\mathrm{bar}$, with maximum local wind speed differences increasing with decreasing pressure.}
    \label{fig:winds_4g}
\end{figure*}
As shown by the streamlines in \Fig{fig:tempmaps_4g}, the wind pattern at pressures $\gtrsim 100~\mathrm{mbars}$ differs between cases with varying interior assumptions. This is most notable at high pressures ($p \approx 10~\mathrm{bars}$), where the fixed flux case has two westerly mid-latitude jets and the hot interior case is characterized by nearly isotropic flow. Additionally, the location and shape of the Rossby gyres (mid-latitude cyclones and anti-cyclones) at the $1~\mathrm{bar} - 100~\mathrm{mbar}$ level differs between the two cases, as the fixed flux case exhibits a dayside mid-latitude anti-cyclone east of the sub-stellar longitude while the hot interior case has a dayside mid-latitude cyclone west of the sub-stellar longitude. Even at a near-photospere pressure of $28~\mathrm{mbar}$, the longitude of the Rossby wave crest (measured in eddy temperature) lies $9.4^\circ$ further east in the hot interior case relative to the fixed flux case. This demonstrates that the planetary-scale standing wave pattern that both induces and is Doppler shifted by the superrotating equatorial jet \citep{Showman_Polvani_2011,Tsai:2014,Hammond:2018aa,Lewis:2022aa} is affected by the assumed interior heat flux and resulting deep atmospheric structure.

The impact of interior assumptions on the planetary-scale wave pattern that drives the equatorial jet in turn implies that the assumed interior properties can affect atmospheric wind speeds. \Fig{fig:winds_4g} shows the horizontal wind speed (i.e., $\sqrt{u^2 + v^2}$, where $u$ is the zonal wind and $v$ is the meridional wind) on isobars from the fixed flux (left) and hot interior (center) cases, along with their difference (right). As discussed above, the different spatial pattern of the wind speeds at depth between the two cases leads to significant contrasts in wind speed in the mid-latitudes. The difference in wind speed between the two cases increases with decreasing pressure, with the largest wind speed differences found in the core of the superrotating equatorial jet at $p = 1 ~\mathrm{bar} - 100~\mathrm{mbars}$ and on the nightside in regions of mid-latitude cyclonic flow at $p = 10 ~\mathrm{mbars}- 1~\mathrm{mbar}$ that are spatially offset between the two cases.
The average difference in horizontal wind speed at the terminator between the two cases at $1~\mathrm{mbar}$ (roughly the level probed in high-resolution transmission spectroscopy, see \citealp{Kempton:2012vk,showman_2013_doppler,Flowers:2018aa}) is only $208~\mathrm{m}~\mathrm{s}^{-1}$. However, the differences in components of the wind vectors can be larger than the differences in wind speeds -- for instance,
the difference in the limb-averaged zonal winds between the two cases is $935~\mathrm{m}~\mathrm{s}^{-1}$, which is comparable to the characteristic present-day observational uncertainty in ground-based high-resolution transmission spectra \citep{Ehrenreich:2020aa,Kesseli:2021ab}.

We also find a difference in the strength of the equatorial jet between the two cases. For instance, the zonal-mean zonal wind speed at 100 mbars (near the core of the jet) at the equator is $604~\mathrm{m}~\mathrm{s}^{-1}$ larger in the fixed flux case than in the hot interior case, a $15.7\%$ difference. As a result, assumptions for the interior heat flux can influence the jet speed at a comparable level to the effect of numerical dissipation \citep{Heng:2011a,Koll:2017,Hammond:2022aa}. Though we do not conduct a thorough analysis of the effect of interior assumptions on the wave-mean flow interactions that drive the superrotating jet, we speculate that the more uniform interior forcing in the hot interior case leads to the weaker jet speed. In principle, the jet speed is regulated by the amplitude of planetary scale standing waves and resulting eddy-mean flow interactions \citep{Showman_Polvani_2011}, implying that reductions in the day-night forcing will lead to a slower equatorial jet.

\subsection{Consistently varying irradiation and interior heat flux}
\label{sec:irradgrav}
\subsubsection{Low gravity}
\label{sec:lowgravity}
\begin{figure*}
    \centering
    \includegraphics[width=0.845\textwidth]{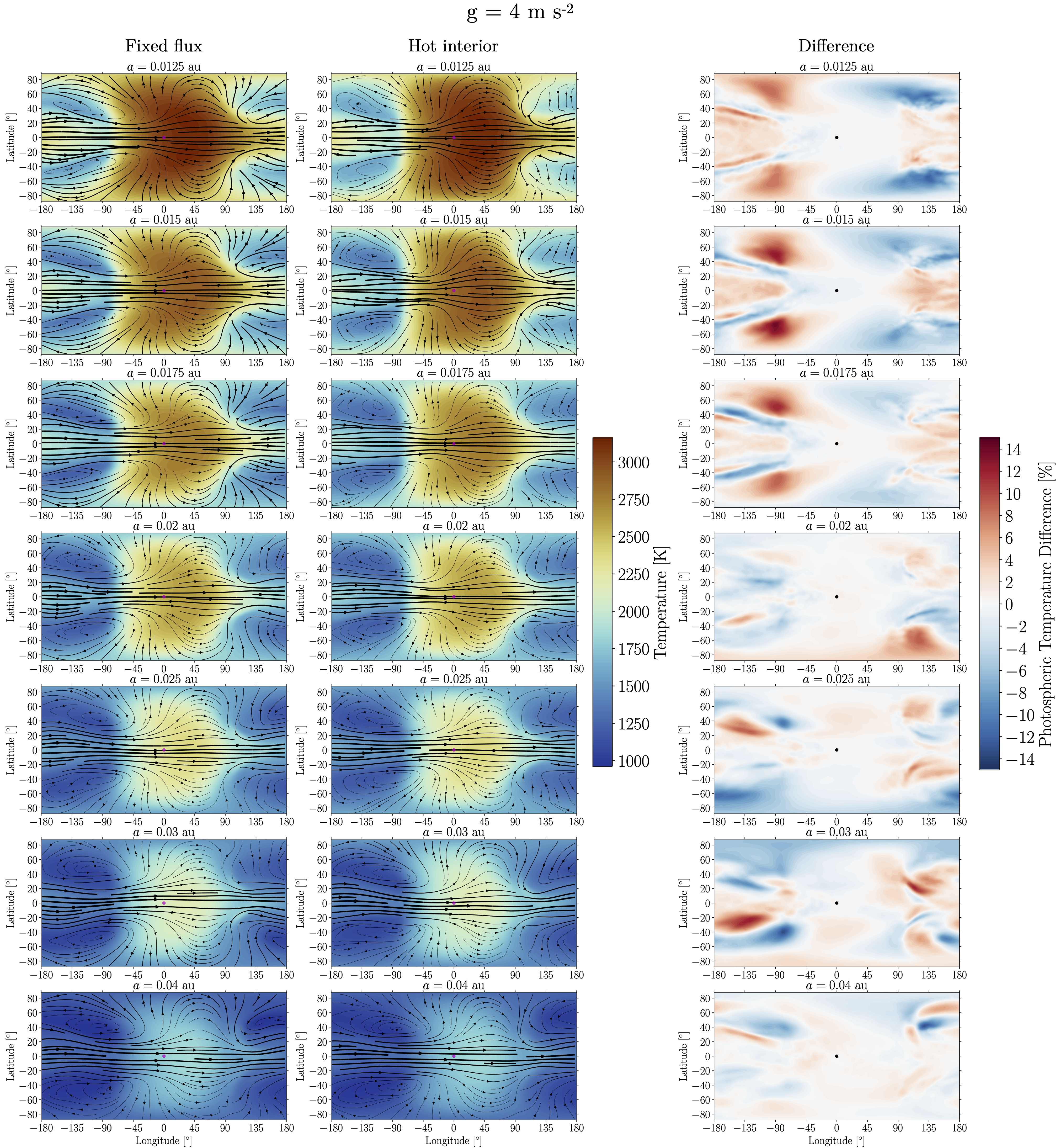}
    \caption{Temperature maps with overlaid wind streamlines at 28 mbar for cases with a surface gravity of 4 m s$^{-2}$ and varying semi-major axis $a = 0.0125~\mathrm{au} - 0.04~\mathrm{au}$, along with the local percent difference in temperature between the fixed flux and hot interior cases. Streamline width scales with wind speed, and all panels in the two left-most columns and in the rightmost-column, respectively, share a color scale. Local near-photospheric temperature differences of up to $\sim 10\%$ occur in all simulations, with the largest differences occurring in mid-latitudes on the nightside.}
    \label{fig:tempmaps_phot_4g}
\end{figure*}
We find that differences in simulated climate between the fixed flux and hot interior cases extend throughout the parameter space of instellation and surface gravity considered. \Fig{fig:tempmaps_phot_4g} shows the near-photospheric\footnote{The thermal photosphere in the cases with $g = 4~\mathrm{m}~\mathrm{s}^{-2}$ is at $26.3~\mathrm{mbar}$, and the thermal photosphere in the cases with $g = 10~\mathrm{m}~\mathrm{s}^{-2}$ is at $55.6~\mathrm{mbar}$.} temperature and winds from the cases with a low surface gravity of $4~\mathrm{m}~\mathrm{s}^{-2}$ and varying semi-major axis from 0.0125 au - 0.04 au. 
All cases are characterized by a significant day-night temperature contrast that drives a planetary-scale wave pattern and concomitant eastward equatorial jet and hot spot offset \citep{Pierrehumbert:2019vk,Showman:2020rev}. However, local differences  in temperature between the fixed flux and hot interior cases of up to $10\%$ occur for all semi-major axes considered. These local temperature differences generally peak in the mid-latitude nightside and are due to the relative positions of the troughs of the Rossby component of the Matsuno-Gill pattern. Additionally, the fixed flux cases with a small semi-major axis $a \le 0.0175$ au are hotter on the equatorial nightside, linked to the faster equatorial jet discussed above. 

We find that in the low gravity cases, global metrics are less dependent on interior assumptions than local climate properties. Regardless of interior assumptions, we find that the fractional day-night temperature contrast ($\Delta T = 1 - T_\mathrm{night}/T_\mathrm{day}$) slightly increases between the 0.0125 au and 0.02 au cases due to the weakening effects of hydrogen dissociation and recombination \citep{Tan:2019aa}, and then decreases toward cooler planets as expected from \cite{perna_2012,Perez-Becker:2013fv}, and \cite{Komacek:2015}. The maximum percent difference in fractional day-night temperature contrast between the two cases for any semi-major axis with $g = 4~\mathrm{m}~\mathrm{s}^{-2}$ is $2.0\%$, implying that photospheric heat transport is not greatly impacted by interior assumptions. Additionally, the largest difference in the longitudinal shift of the equatorial photospheric temperature maximum from the substellar point between the fixed flux and hot interior case in the cases with $g = 4~\mathrm{m}~\mathrm{s}^{-2}$ is $-5.6^\circ$ (with the hot interior case having a more eastward offset than the fixed flux case), and in all other cases with varying semi-major axis the magnitude of the difference is $< 4^\circ$. As a result, we expect that the impact of interior assumptions on large-scale emergent properties is limited for low-gravity planets. 
\subsubsection{Higher gravity}
\begin{figure*}[ht!]
    \centering
    \includegraphics[width=0.845\textwidth]{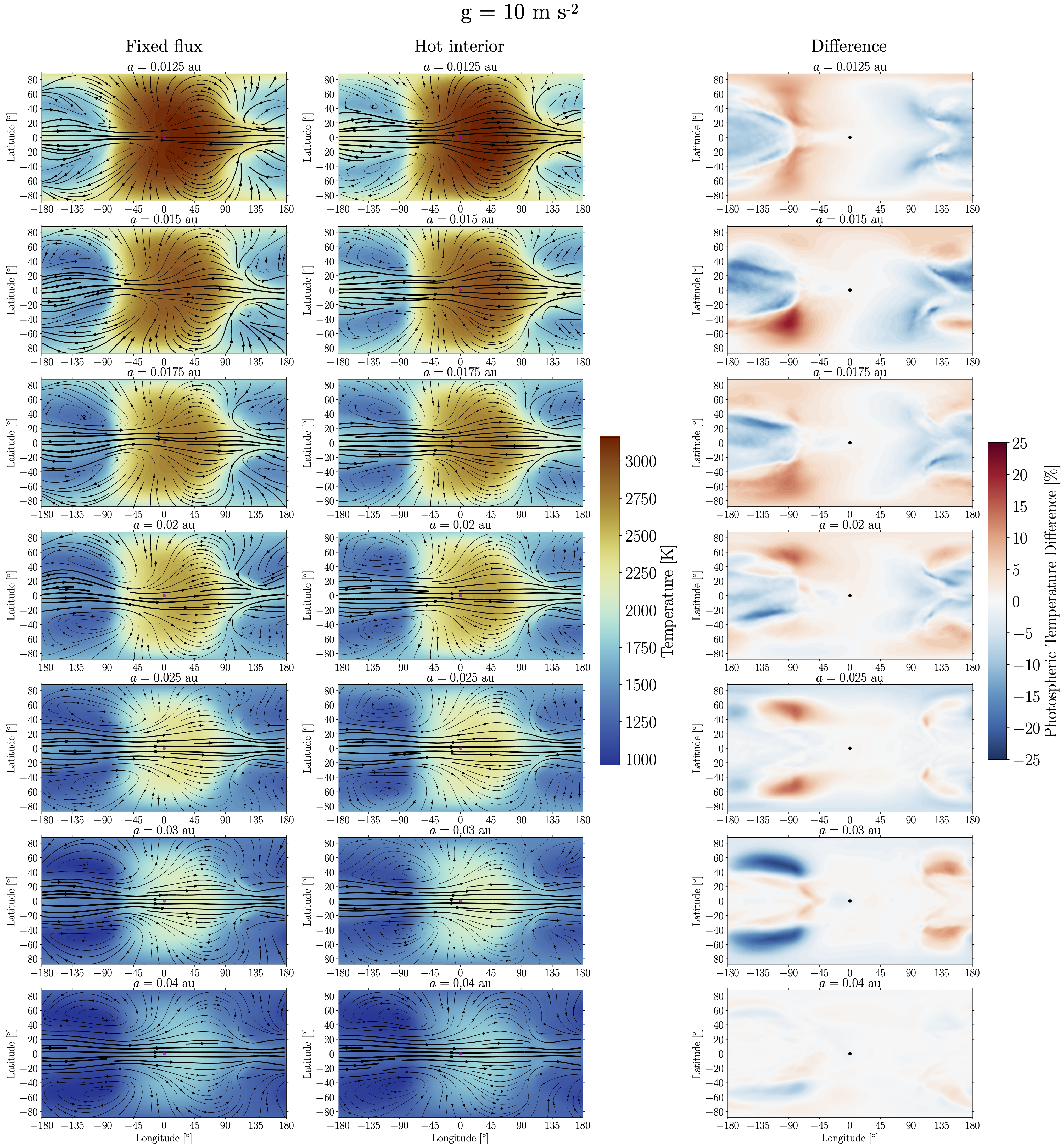}
    \caption{As in \Fig{fig:tempmaps_phot_4g}, but showing temperature maps with overlaid wind streamlines at 57 mbar for cases with a surface gravity of 10 m s$^{-2}$ and varying semi-major axis, along with the local percent difference in temperature between the fixed flux and hot interior cases. Maximum local near-photospheric temperature differences are larger than in the lower-gravity case due to the higher photosphere pressure, and similarly peak in mid-latitudes on the nightside.}
    \label{fig:tempmaps_phot_10g}
\end{figure*}
Given that varying gravity alone only causes a vertical re-scaling of the solution to the dry hydrostatic primitive equations 
\citep{Thomson-2019-Effects}, the greatest impacts of surface gravity on the simulated climate in our GCMs are through the effect on the photosphere pressure and the assumed internal heat flux. As discussed in Sections \ref{sec:parametersweep} and \ref{sec:onecase}, the hot interior cases with a higher gravity have a lower internal heat flux. 
Conversely, a higher gravity increases the pressure at which the atmosphere becomes optically thick for both incoming stellar and outgoing thermal radiation, given an equivalent opacity profile. 

\Fig{fig:tempmaps_phot_10g} shows near-photospheric temperature and wind maps equivalent to those in \Fig{fig:tempmaps_phot_4g} but for the cases with $g = 10~\mathrm{m}~\mathrm{s}^{-2}$. We find that higher gravity strengthens the impact of a hot interior: 
the local differences in near-photospheric temperature between the fixed flux and hot interior cases are larger in the simulations with $g = 10~\mathrm{m}~\mathrm{s}^{-2}$ than in those with $g = 4~\mathrm{m}~\mathrm{s}^{-2}$, and can exceed $20\%$ locally. As in the lower-gravity case, the largest differences generally occur in the mid-latitudes and are due to relative shifts in the location of the mid-latitude Rossby gyres. 

Similarly to the local photospheric temperature, in the higher gravity case we find larger differences between interior assumptions in planetary-scale quantities. The maximum percent difference in the fractional day-night temperature contrast between the fixed flux and hot interior cases occurs in the $a = 0.0125~\mathrm{au}$ case and is $5.1\%$, over twice as large as the maximum difference in the low gravity case. We find that in the higher gravity cases the day-night temperature contrast is always larger in the fixed flux case than in the hot interior cases. 

We also find that the largest difference in the longitudinal offset of the hottest point at the infrared photosphere on the equator is $-11.3^\circ$ (with the hot interior case having a more eastward offset). The hot spot offset is more eastward in the hot interior case than in the fixed flux case for all semi-major axes considered with $g = 10~\mathrm{m}~\mathrm{s}^{-2}$, except the $0.03~\mathrm{au}$ case where the offsets are equivalent. Given that the effect of interior assumptions on climate at observable pressure levels increases with increasing gravity, we suggest that the interior evolution may be especially relevant when making quantitative predictions for the atmospheric circulation of hot and ultra-hot Jupiters that have higher gravities and/or deeper photospheres. 

\subsubsection{Phase curves}

\begin{figure*}
\includegraphics[width=0.495\textwidth]{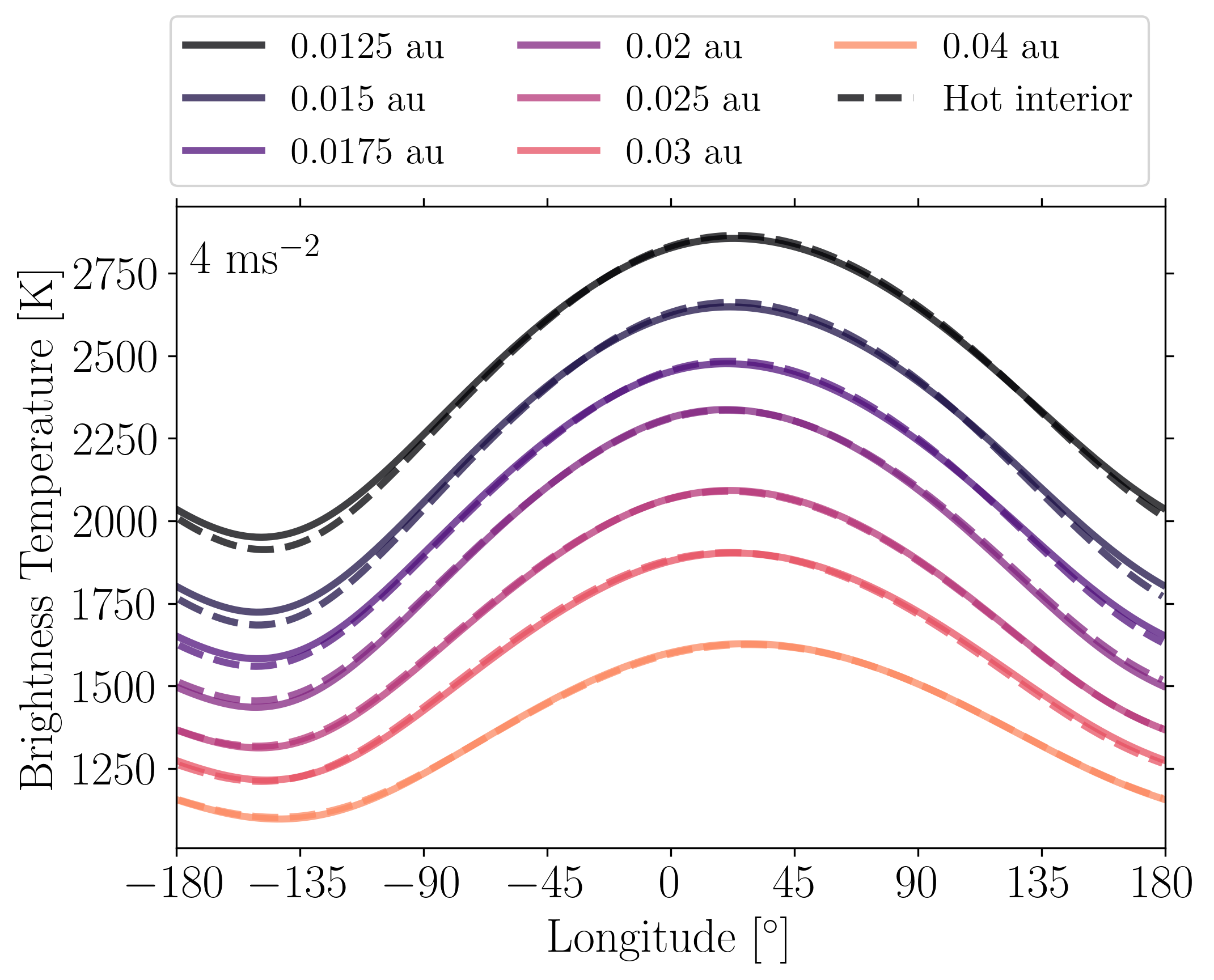} 
\includegraphics[width=0.475\textwidth]{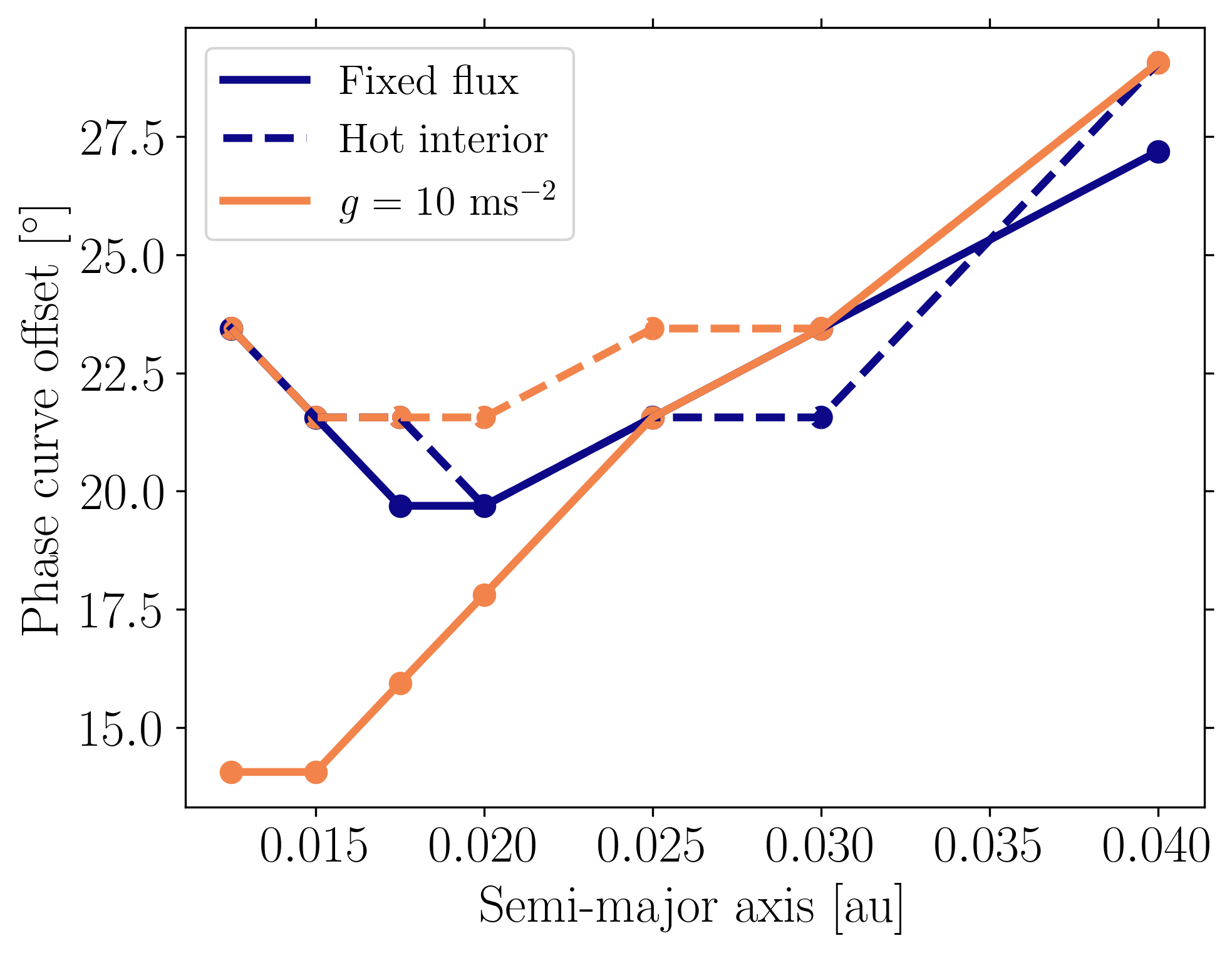}\\
\includegraphics[width=0.495\textwidth]{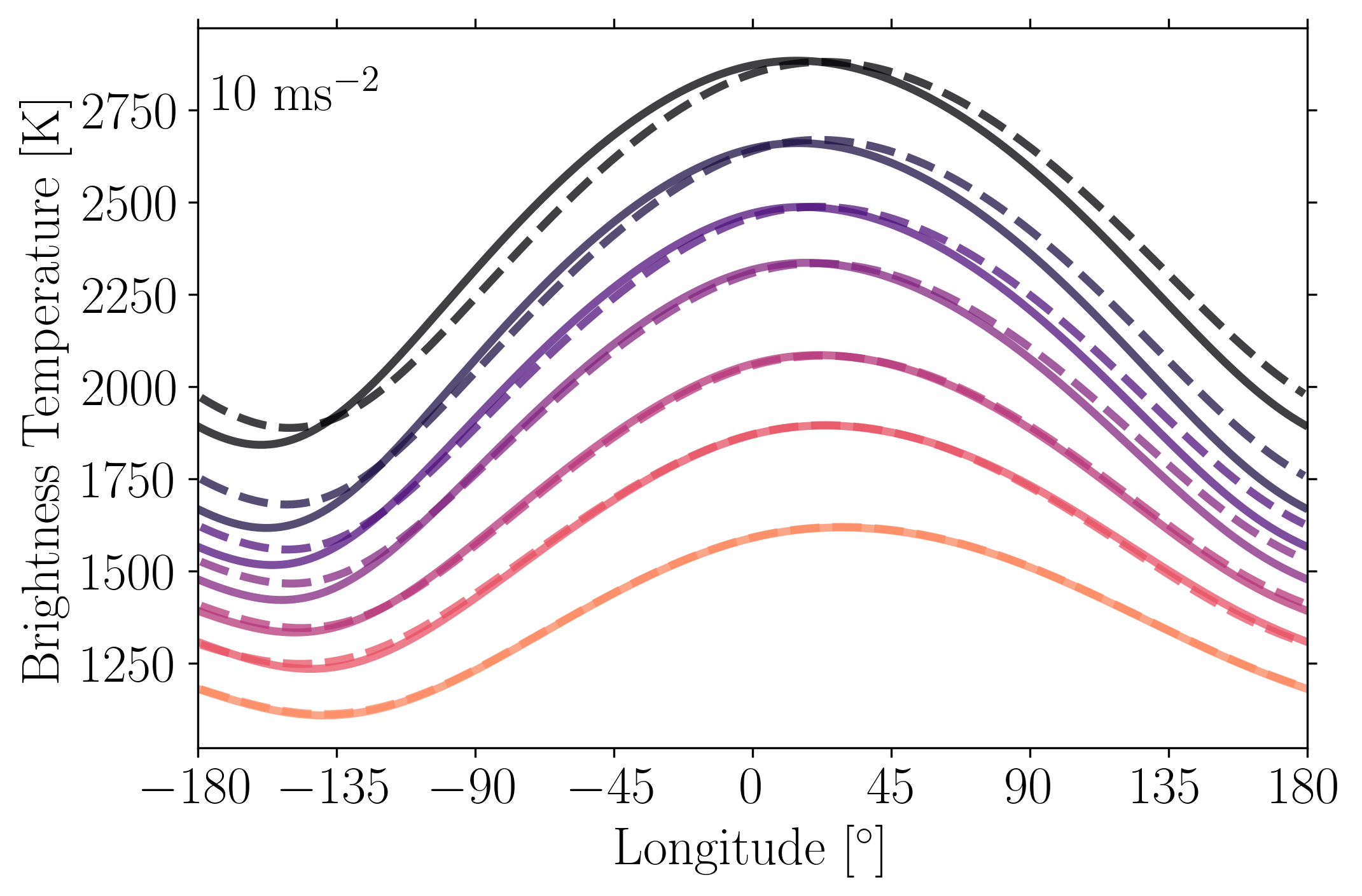}
\includegraphics[width=0.475\textwidth]{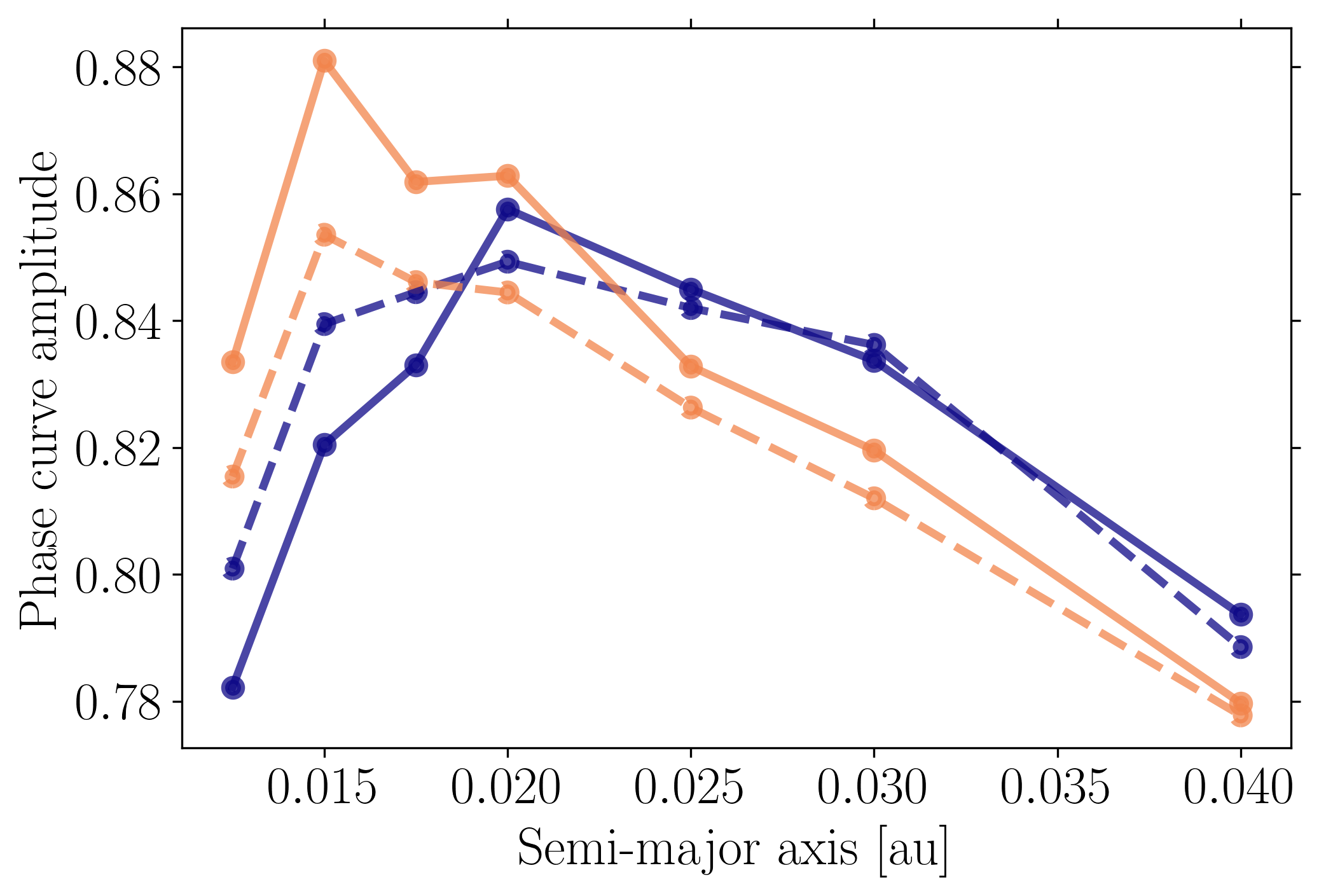}
    \caption{Left: Bolometric infrared brightness temperature as a function of sub-observer longitude with varying semi-major axis (colors, darker lines correspond to smaller separations) and interior assumptions (linestyles, solid lines correspond to fixed flux cases while dashed lines correspond to hot interior cases) for simulations with a surface gravity of $4~\mathrm{m}\mathrm{s}^{-2}$ (top) and $10~\mathrm{m}\mathrm{s}^{-2}$ (bottom).  Right: Phase curve offset (top) and normalized phase curve amplitude  (bottom) as a function of semi-major axis for cases with a fixed flux (solid lines) and hot interior (dashed lines) with a surface gravity of $4~\mathrm{m}\mathrm{s}^{-2}$ (blue) and $10~\mathrm{m}\mathrm{s}^{-2}$ (orange). We find that the effect of interior assumptions on the phase-resolved emission is larger for higher gravity planets.}
    \label{fig:phasecurve}
\end{figure*}

In order to assess the impact of interior assumptions on hemisphere-integrated observable properties, we calculate bolometric thermal phase curves from our full suite of simulations using the method of \cite{Komacek:2017}. The left-hand panels of Figure \ref{fig:phasecurve} show our simulated phase curves in infrared brightness temperature as a function of sub-observer longitude from all simulations in our primary model grid with varying semi-major axis, gravity, and interior assumptions. As in our photospheric temperature maps shown in Figures \ref{fig:tempmaps_phot_4g} and \ref{fig:tempmaps_phot_10g}, we find that the difference between the simulated phase curves of hot interior and fixed flux cases is larger for higher gravity. We also find that the differences between the phase curves calculated with different interior assumptions decrease with increasing semi-major axis. 

 
 From these simulated phase curves, we calculate the phase curve offset as the difference between the sub-observer longitude of the peak hemisphere-averaged thermal flux $F_\mathrm{max}$ and the sub-stellar longitude, with positive values corresponding to an eastward offset. We also calculate the normalized phase curve amplitude as $\left(F_\mathrm{max}-F_\mathrm{min}\right)/F_\mathrm{max}$, where $F_\mathrm{min}$ is the minimum hemisphere-averaged thermal flux flux over all sub-observer longitudes. The resulting phase curve offset and amplitude for each simulation in our grid of models is shown on the right-hand side of Figure \ref{fig:phasecurve}. As discussed in Section 3.2.1, the effects of hydrogen dissociation and recombination cause a non-monotonic trend in the phase curve amplitude and offset for $a < 0.02~\mathrm{au}$. Additionally, the changing vertical structure with planetary parameters and interior assumptions can affect the eddy component of the flow and impact the phase curve offset \citep{Tsai:2014,Lewis:2022aa}. We find that the phase curve offset and amplitude are comparable between the fixed flux and hot interior cases for $g = 4~\mathrm{m}~\mathrm{s}^{-2}$, but the predicted offset and amplitude differ more for $g = 10~\mathrm{m}~\mathrm{s}^{-2}$. Most notably, we find that the difference in phase curve offsets with $g = 10~\mathrm{m}~\mathrm{s}^{-2}$ increase with decreasing semi-major axis, with a maximum difference of $9.4^\circ$ between the hot interior and fixed flux cases. As a result, the effect of interior assumptions on the longitudinal shift of the Matsuno-Gill pattern (discussed in Section 3.1.2) is directly imprinted on the resulting phase curve offset, with more eastward offsets in cases with a hot interior and small semi-major axis.






\section{Discussion \& Conclusions}
\label{sec:conc}
As in the one-dimensional models of \cite{Thorngren:2019aa}, we find that the stratification of the deep atmosphere is greatly reduced in the presence of a hot interior. The deep thermal structure in the case of a hot interior will limit the expected cold trapping of condensates \citep{Spiegel:2009,Parmentier16}, allowing condensible vapor to be mixed to lower pressures. Additionally, the modified deep thermal structure will affect disequilibrium chemistry by changing 
the vertical quench pressure \citep{Smith:1998,Moses:2011,Fortney:2020va}, which will in turn affect the resulting spatial distribution and abundance of disequilibrium species through three-dimensional mixing \citep{Cooper:2006,Steinrueck:2018aa,Drummond:2020aa,Zamyatina:2022aa}. 

Conversely, our hot interior results broadly agree with the expectations of long-baseline GCMs that find that the stratification of the deep atmosphere reduces as it adjusts to a near-adiabat \citep{Sainsbury-Martinez:2019aa,Schneider:2022uu}. As such, though here we isolate the effect of the interior evolution on atmospheric dynamics, these results are potentially congruent with the prediction that downward heat transport by the atmospheric circulation reduces the interior cooling rate of hot and ultra-hot Jupiters, leading to bloated radii \citep{Guillot_2002,showman_2002,Youdin10,Tremblin:2017}. 

JWST is expected to robustly constrain the temperature-pressure profiles of hot Jupiters to at least the $\approx \pm 30~\mathrm{K}$ level through phase curve observations \citep{Bean:2018aa,Rigby:2022vu}. JWST will further constrain the dayside temperature map for a broad range of hot and ultra-hot Jupiters using the secondary eclipse mapping technique \citep{Rauscher07b,Mansfield:2020ab}. Given that the impact of including a hot interior on the local temperature can reach hundreds of Kelvin, it is plausible that interior assumptions will impact the ability of GCMs to match time-series observations of hot and ultra-hot Jupiters. 

Notably, the interior assumption can potentially impact the interpretation of spectroscopic phase curves with JWST. This is because the effect of the interior on thermal structure will be magnified in continuum regions that probe higher pressures on the nightside \citep{Dobbs-Dixon:2017aa}, though nightside clouds might hide this effect \citep{Gao:2021vp,Parmentier:2021tt,Roman:2021wl}. Though the broadband Spitzer phase curve of WASP-76b showed no direct evidence for the impact of the interior on atmospheric circulation \citep{May:2021ab}, \cite{Fortney:2017aa} and \cite{Thorngren:2019aa} describe how spectroscopic JWST observations could probe the internal adiabat in window regions with little water opacity for highly inflated hot Jupiters. Additionally, planets with a hot interior will have a greater CO/CH$_4$ ratio at the quench point \citep{Moses:2011,Venot:2012}. This implies that the resulting disequilibrium chemistry due to mixing will cause carbon in the atmospheres of planets with hot interiors to be predominantly contained in CO rather than CH$_4$ \citep{Thorngren:2019aa}.  As a result, the hypothesis that the interior evolution impacts the atmospheric circulation of hot and ultra-hot Jupiters may be testable with JWST phase curves and secondary eclipse mapping. 

However, we caution that implementation of the expectations from interior evolution models into GCMs with non-gray radiative transfer is required to provide quantitative predictions for thermal structure, as the lack of coupling to wavelength-dependent radiative transfer is a critical limitation of this study. Future work determining the wavelength dependent impact of interior assumptions could guide observing strategies to determine the effect of interior heat flux on observable properties of ultra hot Jupiters. The other key limitation of this work is that the simulations were not conducted over the long ($\sim 10^4-10^5~\mathrm{day}$) baseline required to reach an equilibrated state in the deep atmosphere without the need for a prescribed term in the momentum equation \citep{Liu:2013}. However, unlike previous work studying the deep dynamics of hot and ultra-hot Jupiters, in this work we included the thermodynamic effect of hydrogen dissociation and recombination. This effect can act as a buffer that reduces the impact of interior heating on ultra-hot Jupiters due to the additional energy input required to both warm gas and liberate hydrogen from molecular to atomic form \citep{Roth:2021un}. We propose that further work along the lines of \cite{Mayne:2017,Sainsbury-Martinez:2019aa,Carone:2019aa,Mendonca:2020aa}, and \cite{Schneider:2022aa,Schneider:2022uu} that includes non-gray radiative transfer, hydrogen dissociation and recombination, and incorporates predictions from evolutionary models is required to study how interior evolution impacts the deep atmospheric dynamics of hot and ultra-hot Jupiters. 

In brief, in this work we conducted suites of GCM simulations of hot and ultra-hot Jupiters with and without interior heat fluxes consistent with their inflated radii. We find that these differences in interior assumptions can have significant consequences for the atmospheric circulation and thermal structure of hot and ultra-hot Jupiters. Below we outline our key conclusions. 
\begin{enumerate}
    \item The internal evolution and atmospheric circulation of hot and ultra-hot Jupiters are coupled through the impact of the interior heat flux on atmospheric thermal structure. As a result, including the interior heat flux that corresponds to the inflated radii of hot and ultra-hot Jupiters in GCM simulations affects the predicted temperature pattern and wind speeds throughout the atmosphere.
    \item The global differences in thermal structure between cases with and without a hot interior are largest at pressures of $\gtrsim 1~\mathrm{bar}$, but local differences of hundreds of K can persist to $\sim 1~\mathrm{mbar}$. Local differences in wind speeds can be several hundreds of m s$^{-1}$ or more and increase with decreasing pressure.
   \item The effect of interior assumptions on the thermal structure of hot and ultra-hot Jupiters may impact the interpretation of JWST and high spectral resolution ground-based observations. In our GCMs with a higher gravity, cases with a hot interior generally have a larger hot spot offset and smaller day-night temperature contrast than those without. As a result, we expect that the effect of the internal evolution on observed thermal emission may be important on planets with a deep photosphere, including high gravity and/or low-metallicity objects.
    \item A prediction for the interior heat flux from an evolutionary model consistent with the planetary radius should be used as input for GCMs of hot and ultra-hot Jupiters in order to fully incorporate the impact of internal evolution on atmospheric circulation and thermal structure. 
\end{enumerate}

\software{\\ MITgcm \citep{Adcroft:2004},
\\ IPython \citep{ipython},
\\ Matplotlib \citep{matplotlib},
\\ NumPy \citep{numpy,numpynew}}

\acknowledgments
We thank the referee for their insightful comments and helpful suggestions, which greatly improved this work. We thank Vivien Parmentier for helpful discussions. The authors acknowledge the University of Maryland supercomputing resources (\url{http://hpcc.umd.edu}) made available for conducting the research reported in this paper. This work was completed with resources provided by the University of Chicago Research Computing Center. 

\bibliography{references,References_all,references_paste,References_terrestrial}

\appendix
\begin{figure*}[h!]
    \centering
    \includegraphics[width=0.95\textwidth]{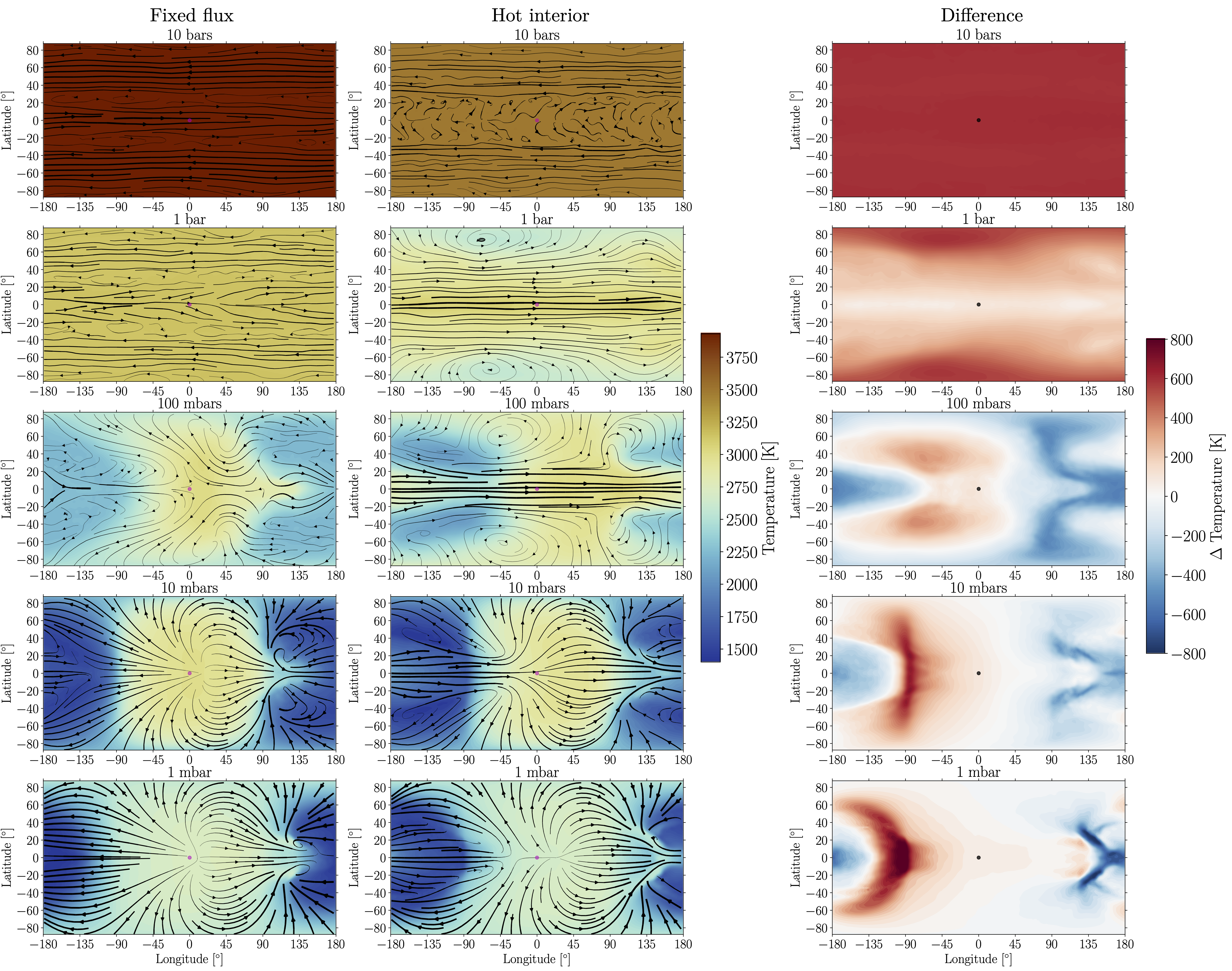}
    \caption{As in Figure \ref{fig:tempmaps_4g}, but from simulations with no additional deep frictional drag applied and $g = 4~\mathrm{m}~\mathrm{s}^{-2}$. Similar to our results including a deep frictional drag, we find that there is a significant difference in the temperature structure between the fixed flux and hot interior cases. Notably, we find that the maximum local differences in temperature between the two interior assumptions can be larger in the cases without a deep frictional drag than in our main suite of GCMs.}
    \label{fig:tempmaps_4g_appendix}
\end{figure*}

\begin{figure*}[h!]
    \centering
    \includegraphics[width=0.95\textwidth]{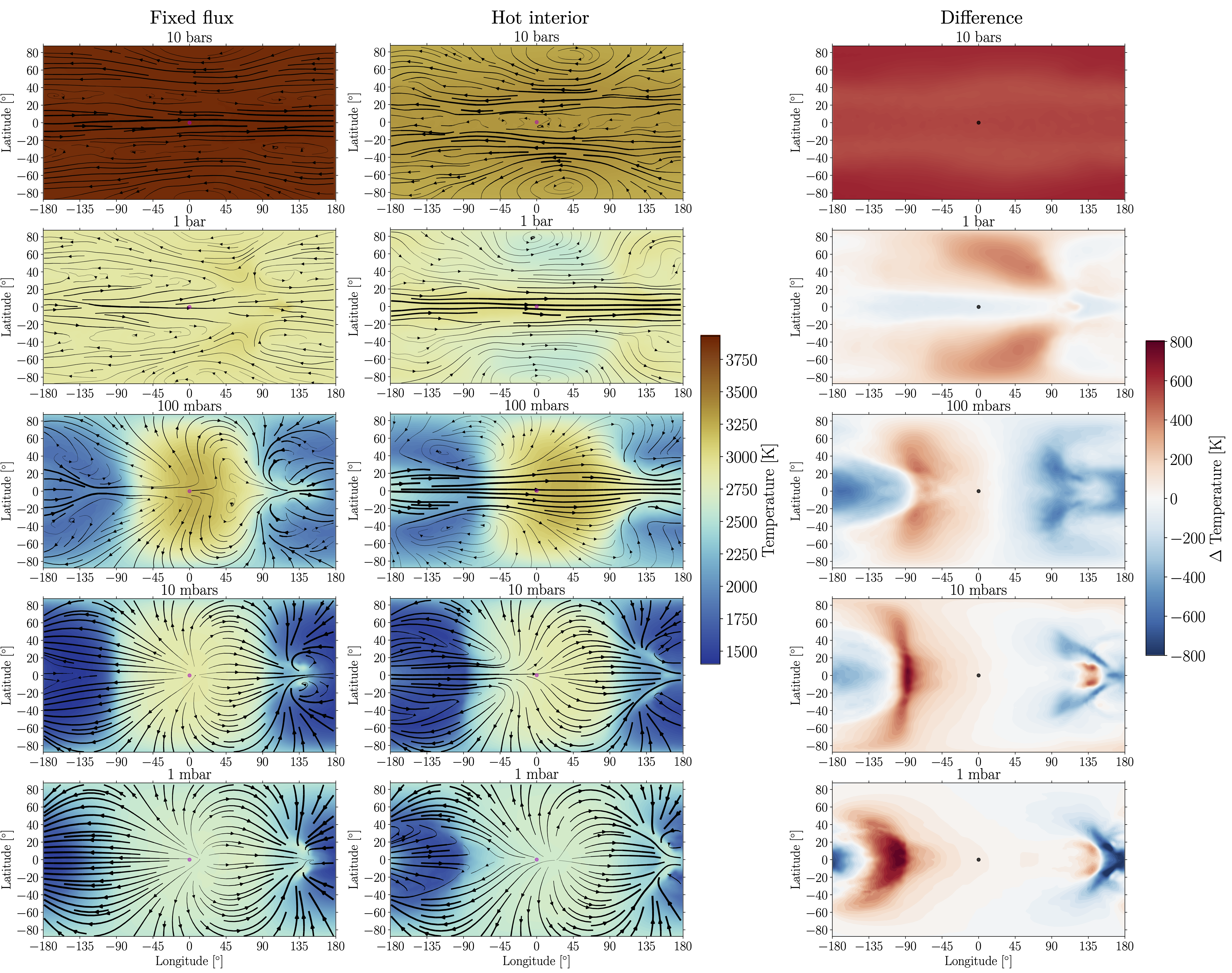}
    \caption{As in Figure \ref{fig:tempmaps_4g}, but from simulations with no additional deep frictional drag applied and $g = 10~\mathrm{m}~\mathrm{s}^{-2}$. Similar to our results for $g = 4~\mathrm{m}~\mathrm{s}^{-2}$ shown in Figure \ref{fig:tempmaps_4g_appendix}, we find that differences between the fixed flux and hot interior assumptions are still significant when not including an additional deep frictional drag.}
    \label{fig:tempmaps_10g_appendix}
\end{figure*}
\section{GCM simulations without deep frictional drag}
\label{app:drag}

In order to investigate the dependence of our results on numerical assumptions, we conducted an additional set of GCM simulations without an applied deep basal drag. Given that our main suite of simulations uses a basal drag that damps the deep circulation at $p > 10~\mathrm{bars}$, it is imperative to determine the extent to which this basal drag assumption affects the results from our main suite of GCMs. Specifically, we our sensitivity tests include four additional simulations covering both the fixed flux and hot interior assumptions and $g = 4~\mathrm{m}~\mathrm{s}^{-2}$ and $10~\mathrm{m}~\mathrm{s}^{-2}$. Besides the removal of the deep basal drag, in order to facilitate direct comparison these models have the same numerical setup and integration time as our main suite of GCMs described in Section \ref{sec:numerics}. 

Figures \ref{fig:tempmaps_4g_appendix} and \ref{fig:tempmaps_10g_appendix} show temperature and wind maps on isobars for the fixed flux and interior cases along with the temperature contrast between the two cases, analogous to Figure \ref{fig:tempmaps_4g} but for cases that do not include an additional frictional drag at the bottom of the domain. Figure  \ref{fig:tempmaps_4g_appendix} is directly comparable to Figure \ref{fig:tempmaps_4g}, as the only difference between the models shown is the presence or absence of additional basal drag. We find that the cases without a deep basal drag show spatial patterns of temperature differences between the fixed flux and hot interior cases that are qualitatively similar to those with a basal drag. Both cases with and without basal drag show that the fixed flux case is globally hotter at $10~\mathrm{bars}$, but at pressures $\lesssim 100~\mathrm{mbar}$ the differences become largest in the Rossby gyres and localized region of adiabatic warming on the nightside near the eastern terminator. Notably, we find that the local differences in temperature at low pressures are larger in cases without a basal drag. Our simulations with $g = 10~\mathrm{m}~\mathrm{s}^{-2}$ and no additional deep drag also show a similar qualitative spatial dependence on the temperature contrasts between fixed flux and hot interior cases with varying pressure level, and show a larger difference in temperature structure when not including the deep frictional drag. 








\end{document}